\def\be{\begin{equation}}
	\def\ee{\end{equation}}
\def\ba{\begin{array}}
	\def\ea{\end{array}}

\documentclass[prl,showpacs,twocolumn,amsmath]{revtex4}
\usepackage{amsmath}
\usepackage{amsfonts}
\usepackage{mathrsfs}
\usepackage{amssymb}
\usepackage{pifont}
\usepackage{epsfig,subfigure,dsfont,amsthm,amsbsy,mathrsfs,amscd}
\usepackage{epstopdf}
\usepackage{bbm}
\usepackage{color}
\def\qed{\leavevmode\unskip\penalty9999 \hbox{}\nobreak\hfil
	\quad\hbox{\leavevmode  \hbox to.77778em{%
			\hfil\vrule   \vbox to.675em%
			{\hrule width.6em\vfil\hrule}\vrule\hfil}}
	\par\vskip3pt}
\usepackage{leftidx}

\input amssym.def
\begin{document}
	\title{\large\bf Quantum state texture of maximal-resource Fourier states under Hawking radiation}
	\author{Shengyan Ma, Yiding Wang, Xiaofen Huang and Tinggui Zhang$^{\dag}$ }
	\affiliation{ School of Mathematics and Statistics, Hainan Normal University, Haikou, 571158, China \\
		\\ $^{\dag}$ Correspondence to tinggui333@163.com}
	
	\bigskip
	\bigskip
	
	\begin{abstract}
	Quantum state texture (QST) is a recently proposed quantum resource that characterizes the nonuniformity of a density matrix in a given computational basis. In the resource theory of QST, the Fourier states $|f_k\rangle$ ($k\ge2$) are maximally resourceful states, attaining the maximum QST measure. We study the QST behavior of three maximally resourceful states $|f_2\rangle$, $|f_3\rangle$, and $|f_4\rangle$ in a two-mode Dirac field under Hawking radiation from a Schwarzschild black hole, using a fidelity-based measure. We consider three cases: only mode $B$ undergoes radiation, only mode $A$ undergoes radiation, and both modes undergo radiation simultaneously. Our results show that for single-mode radiation, if the mode unaffected by radiation is orthogonal to the single-mode textureless state, the QST of the overall output state remains maximal and is independent of the Hawking temperature; we call this phenomenon orthogonality protection. When both modes undergo radiation simultaneously, orthogonality protection vanishes, and the QST of all three Fourier states decreases with increasing Hawking temperature. Notably, the QST measure of $|f_3\rangle$ under two-mode radiation is higher than that under only mode $B$ radiation, indicating that QST can be redistributed among modes: QST gained by one mode under radiation can partially compensate for QST lost by the other. We further provide a general resource-theoretic explanation of orthogonality protection, showing that this phenomenon is independent of the specific form of the QST measure. These results reveal how QST degrades under Hawking radiation and provide a new perspective on the redistribution of quantum resources in black hole backgrounds.
	\end{abstract}
	
	 \maketitle
	
	\section{I. Introduction}
	Quantum state texture (QST) \cite{fp} is a recently proposed quantum resource that characterizes the nonuniformity of the element distribution of a density matrix in a given computational basis. Fixing a computational basis, any density operator can be represented as a table, and QST measures the “roughness” or “irregularity” of the distribution of entries in this table. Unlike quantum coherence, which is usually quantified by the contribution of off-diagonal elements of the density matrix, QST better characterizes the overall distribution of all elements of the density matrix and is therefore more sensitive to phase cancellation and nonuniformity in the element distribution. In this framework, the most uniformly distributed state is called the texture-free state, which is the zero-resource state of QST resource theory. In contrast to the zero-resource state are maximally resourceful states. Reference \cite{fp} shows that these maximally resourceful states are formed by Fourier states, which are orthogonal to the texture-free state and possess the maximum QST resource in resource theory. For the quantification and characterization of QST, subsequent works have proposed fidelity-based measures, trace-distance measures, geometric measures, and various constructions based on relative entropy and Tsallis relative entropy \cite{wyd,czzg,yczf,wmcz,xycq,yjcy,cyzz,ylyg}. Furthermore, QST has been widely studied in quantum phase transitions, open-system dynamics, gate identification, as well as its connections with other quantum resources \cite{aptk,rm,acbg,ywxh,lcck,chsv,yxyc,jadl,whwz}.

	Research at the intersection of quantum information and gravity is a major direction in contemporary theoretical physics. In curved spacetime, the presence of a horizon induces mode mixing in quantum fields, causing the quantum state seen by an external observer to decohere \cite{emlj,jj,jwqp}. Hawking radiation is a paradigmatic manifestation of this effect: vacuum fluctuations near the black hole horizon lead to thermal particle emission, and an external observer can access only a subset of the field modes, while the remaining modes fall inside the horizon and are in principle unobservable \cite{tdrr}. The degradation of various quantum resources, including entanglement, coherence, steering and nonlocality, induced by this mechanism has been extensively studied \cite{ifsr,debj,jwjj,emmi,deba,jcyk,tljj,smwh,smwh2,tzxw,smwx,gwmx,gwmx2,shss,smwh3,gwmx3}. In recent years, QST has also begun to be studied in curved spacetime and non-inertial frames \cite{zhlz,zmh,azyz}. The authors of Ref. \cite{zhlz} showed that for two accelerated atoms coupled to a massive scalar field, QST is dominated by the competition between local decoherence and collective thermalization, approaching an Unruh thermal state at long times. Huang \cite{zmh} found that in Schwarzschild–de Sitter spacetime, the double horizon redistributes QST between accessible and inaccessible modes, with a nonlinear constraint rather than a conservation law. Zhang et al. \cite{azyz} revealed a trade-off between the accessible and inaccessible QST of multipartite GHZ states in non-inertial frames and gave a quantitative bound on its relation to genuine multipartite entanglement. However, these works mainly focus on general states or specific quantum states, and the behavior of maximally resourceful states in QST resource theory in curved spacetime has not been systematically studied. Motivated by this gap, this work studies the QST behavior of three maximally resourceful Fourier states in a two-mode Dirac field under Hawking radiation from a Schwarzschild black hole. This study is important for three reasons. First, maximally resourceful states are central objects in QST resource theory, yet their behavior in curved spacetime has not been systematically studied; investigating their degradation provides a concrete setting for examining the theory beyond flat spacetime. Second, comparing single-mode and two-mode radiation allows us to examine whether the distribution of QST in a composite system depends on which modes undergo radiation, a question not addressed in previous works. Third, analyzing the degradation mechanism within the general framework of QST resource theory can clarify how QST is distributed among subsystems and whether this distribution obeys general constraints.
	
	In this work, we consider three maximally resourceful Fourier states in a two-mode Dirac field. Using a fidelity-based QST measure, we calculate analytical results and numerical plots for three cases: only mode $B$ undergoes radiation, only mode $A$ undergoes radiation, and both modes undergo radiation simultaneously. We find that when only one mode undergoes radiation, if the unaffected mode is orthogonal to the single-mode texture-free state, the overlap between the overall output state and the texture-free state is always zero, and the QST measure remains maximal, independent of the Hawking temperature. We call this phenomenon orthogonality protection. When both modes undergo radiation simultaneously, orthogonality protection vanishes completely, and the QST measures of all three Fourier states decrease with increasing Hawking temperature. In particular, the QST measure of one Fourier state under two-mode radiation is higher than its value under only mode $B$ radiation, indicating that QST can be redistributed among different modes. Finally, starting from the general framework of QST resource theory, we explain the origin of orthogonality protection and show that this phenomenon does not depend on the choice of specific measure. The rest of this paper is organized as follows. Section II introduces the quantization of the Dirac field in Schwarzschild spacetime, QST resource theory, and the definition of Fourier states. Section III presents detailed calculations and resource-theoretic explanations for the three radiation cases. We conclude and discuss future research directions in the final section.
    \section{II. Preliminaries}
    We consider a Schwarzschild black hole of mass $M$, the metric of which can be written as \cite{emlj}:
    \begin{equation*}
    	\begin{split}
    		ds^2=&-(1-\frac{2M}{r})dt^2+(1-\frac{2M}{r})^{-1}dr^2\\
    		&+r^2(d\theta^2+\sin^2\theta d\varphi^2),
    	\end{split}
    \end{equation*}
    The first term on the right-hand side describes how the geometry of the temporal dimension is influenced by mass and distance. The second term describes the curvature of the radial spatial geometry, and the final term represents the geometry of the angular components. Here, $r$ is the radial coordinate, $\theta$ is the polar angle, and $\varphi$ is the azimuthal angle. To simplify the equations, we use natural units by setting the gravitational constant $G$, the speed of light $c$, the Planck constant $\hbar$, and the Boltzmann constant $k_B$ to unity, i.e., $G=c=\hbar=k_B=1$. The equation of motion for a massless Dirac field in Schwarzschild spacetime is given by
    \begin{equation*}
    	\begin{split}
    		&-\frac{\gamma_0}{\sqrt{1-\frac{2M}{r}}}\frac{\partial\Phi}{\partial t}+\gamma_1\sqrt{1-\frac{2M}{r}}[\frac{\partial}{\partial r}+\frac{1}{r}+\frac{M}{2r(r-2M)}]\Phi\\
    		&+\frac{\gamma_2}{r}(\frac{\partial}{\partial\theta}+\frac{\cot\theta}{2})\Phi+\frac{\gamma_3}{r\sin\theta}\frac{\partial\Phi}{\partial\varphi}=0,
    	\end{split}
    \end{equation*}
    where $\gamma_i (i=0,1,2,3)$ are the Dirac matrices \cite{jj,jwqp}. Solving this equation yields a set of positive-frequency solutions, composed of outgoing and ingoing modes \cite{emlj,jj}:
    \begin{align}
    	\Phi_{k,\text{out}}^+&\sim \phi(r)\exp(-i\omega u),\label{e1}\\
    	\Phi_{k,\text{in}}^+&\sim \phi(r)\exp(i\omega u),\label{e2}
    \end{align}
    where $k$ represents the wave vector, $\phi(r)$ is the four component Dirac spinor, $\omega$ denotes a monochromatic frequency, the retarded time defined as $u=t-r_*$, and the tortoise coordinate given by $r_*=r+2M\ln\frac{r-2M}{2M}$. 
    
    A complete basis for the Kruskal modes has been presented in \cite{tdrr}. By quantizing the Dirac field in the Schwarzschild and Kruskal coordinates separately, the Bogoliubov transformation connecting the creation and annihilation operators in these coordinate systems is derived. Accordingly, the Kruskal vacuum and excited states in Schwarzschild spacetime can be expressed as \cite{smwh,smwx}
    \begin{small}
    	\begin{equation}\label{e5}
    		\begin{split}
    			|0\rangle_K&=\frac{1}{\sqrt{e^{-\frac{\omega}{T}}+1}}|0\rangle_\text{out}|0\rangle_\text{in}+\frac{1}{\sqrt{e^{\frac{\omega}{T}}+1}}|1\rangle_\text{out}|1\rangle_\text{in},\\
    			|1\rangle_K&=|1\rangle_\text{out}|0\rangle_\text{in}.
    		\end{split}
    	\end{equation}
    \end{small}
    These relations imply that, to a distant inertial observer using Schwarzschild coordinates, the vacuum outside the black hole is not a pure vacuum state but an entangled one. Here, $T=\frac{1}{8\pi M}$ is the Hawking temperature.
    
    Now, we introduce the concept of quantum-state texture (QST) \cite{fp}. Fix a $D$-dimensional computational basis $\{|i\rangle\}$. Any density operator $\rho$ can be viewed as a $D\times D$ table with entries $\rho_{ij}=\langle i|\rho|j\rangle$. QST characterizes the irregularity of the distribution of entries in this table. In this framework, the most uniform state is:
    \begin{equation}
    |f_1\rangle=\frac{1}{\sqrt{D}}\sum_{i=0}^{D-1}|i\rangle,\,f_1=|f_1\rangle\langle f_1|.
    \end{equation}
    All its matrix elements are equal in the computational basis. This state is called the texture-free state and is the zero-resource state of the resource theory. Following Ref. \cite{fp}, the Fourier states in QST are defined as
    \begin{equation}
    	|f_k\rangle = \frac{1}{\sqrt{D}} \sum_{j=0}^{D-1} \omega_D^{(k-1)j} |j\rangle,
    \end{equation}
    with $\omega_D = e^{2\pi i/D}$ and $k = 1,\dots,D$. For $k=1$, we recover the textureless state $|f_1\rangle$. For $k\geq2$, one has $\langle f_1|f_k\rangle=0$, so that $f_k$ lies in the orthogonal support of $f_1$. It has been shown in \cite{fp} that these Fourier states are the maximal-resource states in the QST resource theory.
    
    To quantify the texture of an arbitrary state $\rho$, we adopt the fidelity-based measure proposed in Ref.\cite{wyd}
    \begin{equation}
    	\mathcal{T}_F(\rho)=1-F(\rho,f_1)=1-\frac{\Sigma(\rho)}{D},
    \end{equation}
    where $F$ is the Uhlmann fidelity and $\Sigma(\rho)=\sum_{i,j}\rho_{ij}$ is the grand sum defined in \cite{fp}. This measure satisfies the basic requirements for a QST measure and attains its maximum for the maximally resourceful Fourier state: $\mathcal{T}_F(f_k)=1, k\geq2$. Therefore, the fidelity-based measure is bounded and suitable for describing the behavior of maximally textured states under Hawking radiation.
    
    Next, we take $|f_k\rangle$ ($k\geq 2$) as the initial state. After a Bogoliubov transformation in Schwarzschild spacetime, we trace out the inaccessible modes to obtain the accessible output state $\rho_\text{out}^{(k)}$ and compute its fidelity-based measure $\mathcal{T}_F(\rho_\text{out}^{(k)})$. We further define the texture distinguishability between two output states as the difference in their fidelity-based measures:
    \begin{equation}
    	\mathcal{D}_{kj}=|\mathcal{T}_F(\rho_\text{out}^{(k)})-\mathcal{T}_F(\rho_\text{out}^{(j)})|.
    \end{equation}
    This work analyzes how the Hawking temperature $T=\frac{1}{8\pi M}$ affects QST and the texture distinguishability of these maximally resourceful states.
    \section{III. Behavior of Fourier states in a two-mode Dirac field under Hawking radiation}
    In this subsection, we consider two Dirac field modes, $A$ and $B$, in Schwarzschild spacetime. The occupation-number degree of freedom of each mode forms a two-level system, i.e., a qubit. Hence, the total Hilbert space dimension is $D=4$, and the general Fourier states reduce to
    \begin{equation}
    	|f_k\rangle = \frac{1}{2} \sum_{j=0}^{3} \omega_4^{(k-1)j} |j\rangle,\,\omega_4=e^{2\pi i/4}=i,
    \end{equation}
    where the single-index basis $\{|0\rangle,|1\rangle,|2\rangle,|3\rangle\}$ is identified with the two-mode basis $\{|00\rangle,|01\rangle,|10\rangle,|11\rangle\}$ via $|0\rangle\to|00\rangle$, $|1\rangle\to|01\rangle$, $|2\rangle\to|10\rangle$, $|3\rangle\to|11\rangle$.
     Here the first index corresponds to mode $A$ and the second to mode $B$. 
    
    \subsection{Only Mode $B$ Undergoes Hawking Radiation}
    Assume that only mode $B$ undergoes Hawking radiation from a Schwarzschild black hole, while mode $A$ is unaffected. The relation between the Kruskal state of mode $B$ and the exterior and interior modes is
    \begin{equation*}
    		\begin{split}
    			|0\rangle_K^B&=\mu|0\rangle_\text{out}^B|0\rangle_\text{in}^B+\nu|1\rangle_\text{out}^B|1\rangle_\text{in}^B,\\
    			|1\rangle_K^B&=|1\rangle_\text{out}^B|0\rangle_\text{in}^B,
    		\end{split}
    	\end{equation*}
      where the Bogoliubov coefficients are
      \begin{equation*}
      \mu=\frac{1}{\sqrt{e^{-\frac{\omega}{T}}+1}},\,\nu=\frac{1}{\sqrt{e^{\frac{\omega}{T}}+1}},\,\mu^2+\nu^2=1.
      \end{equation*}
    The Hawking temperature is $T=\frac{1}{2\pi M}$. In the numerical calculations, we take the mode frequency $\omega=\pi$ for concreteness. Clearly, as $T\to 0$, $\mu\to 1$ and $\nu\to 0$; as $T\to\infty$, $\mu\to \frac{1}{\sqrt{2}}$ and $\nu\to \frac{1}{\sqrt{2}}$. In the computational basis, the Bogoliubov transformation acts as:
    \begin{equation*}
    \begin{split}
    &|0\rangle^A|0\rangle^B\to\mu|0\rangle^A|0\rangle_\text{out}^B|0\rangle_\text{in}^B+\nu|0\rangle^A|1\rangle_\text{out}^B|1\rangle_\text{in}^B,\\
    &|0\rangle^A|1\rangle^B\to|0\rangle^A|1\rangle_\text{out}^B|0\rangle_\text{in}^B,\\
    &|1\rangle^A|0\rangle^B\to\mu|1\rangle^A|0\rangle_\text{out}^B|0\rangle_\text{in}^B+\nu|1\rangle^A|1\rangle_\text{out}^B|1\rangle_\text{in}^B,\\
    &|1\rangle^A|1\rangle^B\to|1\rangle^A|1\rangle_\text{out}^B|0\rangle_\text{in}^B.
    \end{split}
    \end{equation*}    
    For notational convenience, in the following we abbreviate the out-mode ket as $|ij\rangle\equiv|i\rangle^A|j\rangle_\text{out}^B$ and omit the label $B$ on the interior mode.
    
    From Eq. (8), the maximal textured states for the two Dirac field modes are:
    \begin{equation*}
    \begin{split}
    |f_2\rangle&=\frac{1}{2}(|00\rangle+i|01\rangle-|10\rangle-i|11\rangle),\\
    |f_3\rangle&=\frac{1}{2}(|00\rangle-|01\rangle+|10\rangle-|11\rangle),\\
    |f_4\rangle&=\frac{1}{2}(|00\rangle-i|01\rangle-|10\rangle+i|11\rangle).
    \end{split}
    \end{equation*}
    In the following, we denote by $|\Psi_k\rangle$ the joint state of the exterior and interior modes obtained after applying the Bogoliubov transformation to $|f_k\rangle$ in mode $B$. Taking the partial trace over the interior mode yields the externally accessible output state:
    \begin{equation}
    \rho_\text{out}^{(k)}=\text{Tr}_\text{in}(|\Psi_k\rangle\langle\Psi_k|).
    \end{equation}
    Next, we perform detailed calculations for the three maximally textured states.
    
    {\boldmath $|f_2\rangle$:} Substituting the Bogoliubov transformation into the expression for $|f_2\rangle$, we obtain:
    \begin{equation*}
    \begin{split}
    |\Psi_2\rangle&=\frac{1}{2}[(\mu|00\rangle+i|01\rangle-\mu|10\rangle-i|11\rangle)|0\rangle_\text{in}\\
    &+\nu(|01\rangle-|11\rangle)|1\rangle_\text{in}]\\
    &=\frac{1}{2}(|\psi_0^{(2)}\rangle|0\rangle_\text{in}+|\psi_1^{(2)}\rangle|1\rangle_\text{in}),
    \end{split}
    \end{equation*}
    where 
    \begin{equation*}
    \begin{split}
    &|\psi_0^{(2)}\rangle=\mu|00\rangle+i|01\rangle-\mu|10\rangle-i|11\rangle,\\
    &|\psi_1^{(2)}\rangle=\nu(|01\rangle-|11\rangle).
    \end{split}
    \end{equation*}
    Tracing out the interior mode gives:
    \begin{equation*}
    \rho_\text{out}^{(2)}=\text{Tr}_\text{in}(|\Psi_2\rangle\langle\Psi_2|)=\frac{1}{4}(|\psi_0^{(2)}\rangle\langle\psi_0^{(2)}|+|\psi_1^{(2)}\rangle\langle\psi_1^{(2)}|).
    \end{equation*}
    A direct calculation shows that:
    \begin{equation*}
    \langle f_1|\psi_0^{(2)}\rangle=\langle f_1|\psi_1^{(2)}\rangle=0.
    \end{equation*}
    Thus one has
    \begin{equation*}
    \mathcal{T}_F(\rho_\text{out}^{(2)})=1-\langle f_1|\rho_\text{out}^{(2)}|f_1\rangle=1.
    \end{equation*}
    Note that the equality above follows from $f_1$ being a pure state, so $F(\rho_\text{out}^{(2)},f_1)=\langle f_1|\rho_\text{out}^{(2)}|f_1\rangle$.
    
    {\boldmath $|f_3\rangle$:} Substituting the Bogoliubov transformation into the expression for $|f_3\rangle$, one obtains:
    \begin{equation*}
    	\begin{split}
    		|\Psi_3\rangle&=\frac{1}{2}[(\mu|00\rangle-|01\rangle+\mu|10\rangle-|11\rangle)|0\rangle_\text{in}\\
    		&+\nu(|01\rangle+|11\rangle)|1\rangle_\text{in}]\\
    		&=\frac{1}{2}(|\psi_0^{(3)}\rangle|0\rangle_\text{in}+|\psi_1^{(3)}\rangle|1\rangle_\text{in}),
    	\end{split}
    \end{equation*}
    where 
    \begin{equation*}
    	\begin{split}
    		&|\psi_0^{(3)}\rangle=\mu|00\rangle-|01\rangle+\mu|10\rangle-|11\rangle,\\
    		&|\psi_1^{(3)}\rangle=\nu(|01\rangle+|11\rangle).
    	\end{split}
    \end{equation*}
    Tracing out the interior mode yields:
    \begin{equation*}
    	\rho_\text{out}^{(3)}=\text{Tr}_\text{in}(|\Psi_3\rangle\langle\Psi_3|)=\frac{1}{4}(|\psi_0^{(3)}\rangle\langle\psi_0^{(3)}|+|\psi_1^{(3)}\rangle\langle\psi_1^{(3)}|).
    \end{equation*}
    Next, we compute the projection onto the textureless state $|f_1\rangle$:
    \begin{equation*}
    \begin{split}
    \langle f_1|\rho_\text{out}^{(3)}|f_1\rangle&=\frac{1}{4}(|\langle f_1|\psi_0^{(3)}\rangle|^2+|\langle f_1|\psi_1^{(3)}\rangle|^2)\\
    &=\frac{1}{4}(\mu^2-2\mu+1+\nu^2)\\
    &=\frac{1-\mu}{2}.
    \end{split}
    \end{equation*}
    Therefore, the fidelity-based measure is
    \begin{equation*}
    \mathcal{T}_F(\rho_\text{out}^{(3)})=1-\langle f_1|\rho_\text{out}^{(3)}|f_1\rangle=\frac{1+\mu}{2}.
    \end{equation*}
    
    {\boldmath $|f_4\rangle$:} Similarly, for $|f_4\rangle$, we have:
    \begin{equation*}
    	\begin{split}
    		|\Psi_4\rangle&=\frac{1}{2}[(\mu|00\rangle-i|01\rangle-\mu|10\rangle+i|11\rangle)|0\rangle_\text{in}\\
    		&+\nu(|01\rangle-|11\rangle)|1\rangle_\text{in}]\\
    		&=\frac{1}{2}(|\psi_0^{(4)}\rangle|0\rangle_\text{in}+|\psi_1^{(4)}\rangle|1\rangle_\text{in}),
    	\end{split}
    \end{equation*}
    with
    \begin{equation*}
    	\begin{split}
    		&|\psi_0^{(4)}\rangle=\mu|00\rangle-i|01\rangle-\mu|10\rangle+i|11\rangle,\\
    		&|\psi_1^{(4)}\rangle=\nu(|01\rangle-|11\rangle).
    	\end{split}
    \end{equation*}
    According to Eq. (9), we can obtain that
    \begin{equation*}
  	\rho_\text{out}^{(4)}=\text{Tr}_\text{in}(|\Psi_4\rangle\langle\Psi_4|)=\frac{1}{4}(|\psi_0^{(4)}\rangle\langle\psi_0^{(4)}|+|\psi_1^{(4)}\rangle\langle\psi_1^{(4)}|).
    \end{equation*}
    It is straightforward to verify that:
    \begin{equation*}
    	\mathcal{T}_F(\rho_\text{out}^{(4)})=1-\langle f_1|\rho_\text{out}^{(4)}|f_1\rangle=1.
    \end{equation*}
    
    Now, we summarize the fidelity-based measures of the three maximally textured resource states:
    \begin{equation*}
    \begin{split}
    &\mathcal{T}_F(\rho_\text{out}^{(2)})=\mathcal{T}_F(\rho_\text{out}^{(4)})=1,\\
    &\mathcal{T}_F(\rho_\text{out}^{(3)})=\frac{1+\mu}{2}.
    \end{split}
    \end{equation*}
    Using the texture distinguishability defined in Eq. (7), we obtain:
    \begin{equation*}
    \mathcal{D}_{24}=0,\,\mathcal{D}_{23}=\mathcal{D}_{34}=\frac{1-\mu}{2}.
    \end{equation*}
    We examine the limiting behavior: as $T\to 0$, $\mu\to 1$, then 
    \begin{equation*}
    \mathcal{T}_F(\rho_\text{out}^{(3)})\to1,\,\mathcal{D}_{23}\to0.
    \end{equation*}
     This means that all maximally resourceful states retain maximal texture and are indistinguishable in the QST measure. However, as $T\to\infty$, $\mu\to \frac{1}{\sqrt{2}}$, then
     \begin{equation*}
     	\mathcal{T}_F(\rho_\text{out}^{(3)})\to\frac{1+\frac{1}{\sqrt{2}}}{2},\,\mathcal{D}_{23}\to\frac{1-\frac{1}{\sqrt{2}}}{2}.
     \end{equation*}
     Figure 1 shows the fidelity-based measure $\mathcal{T}_F(\rho_\text{out}^{(3)})$ and the texture distinguishability $\mathcal{D}_{23}$ ($\mathcal{D}_{34}$) versus the Hawking temperature $T$.
     
     \begin{figure}[htbp]
     	\centering
     	\includegraphics[width=0.5\textwidth]{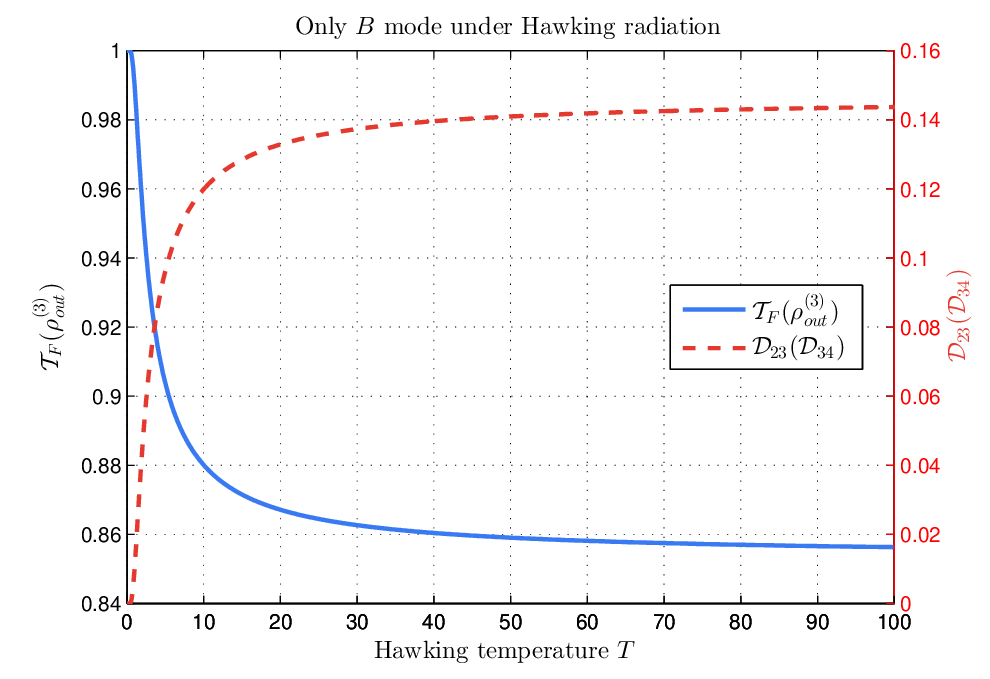}
     	\vspace{-1em} \caption{Variation of the fidelity-based measure $\mathcal{T}_F(\rho_\text{out}^{(3)})$ and the texture distinguishability $\mathcal{D}_{23}$ ($\mathcal{D}_{34}$) with Hawking temperature $T$ when only mode $B$ undergoes Hawking radiation. The blue solid line denotes $\mathcal{T}_F(\rho_\text{out}^{(3)})$, and the red dashed line denotes $\mathcal{D}_{23}$ ($\mathcal{D}_{34}$). $\mathcal{T}_F(\rho_\text{out}^{(3)})$ decreases monotonically from 1 to about 0.854, while $\mathcal{D}_{23}$ ($\mathcal{D}_{34}$) increases monotonically from 0 to about 0.146.} \label{Fig.1}
     \end{figure}
      
      These results show that when only mode $B$ undergoes Hawking radiation, the QST measures of $|f_2\rangle$ and $|f_4\rangle$ remain constant at 1, whereas the QST of $|f_3\rangle$ decreases monotonically with increasing Hawking temperature, and its texture distinguishability from $|f_2\rangle$ and $|f_4\rangle$ increases monotonically. This difference is not accidental; it arises from the different texture distributions of the three maximally resourceful states in the subsystems. To understand this, we introduce the single-mode superposition state:
      \begin{equation}
      |\pm\rangle=\frac{1}{\sqrt{2}}(|0\rangle\pm|1\rangle),\,|\pm i\rangle=\frac{1}{\sqrt{2}}(|0\rangle\pm i|1\rangle),
      \end{equation}
      where $|+\rangle$ and $|-\rangle$ are the single-mode texture-free and maximally textured states, respectively. Using these states, the three maximally resourceful Fourier states can be written in product form:
      \begin{equation*}
      \begin{split}
      &|f_2\rangle=|-\rangle\otimes|+i\rangle,\\
      &|f_3\rangle=|+\rangle\otimes|-\rangle,\\
      &|f_4\rangle=|-\rangle\otimes|-i\rangle.
      \end{split}
      \end{equation*}
      This decomposition implies that, although all three states are maximally resourceful, the distribution of QST between modes $A$ and $B$ is completely different: the mode-$A$ part of $|f_3\rangle$ is exactly the texture-free state, so all QST is stored in mode $B$. In contrast, the mode-$A$ parts of $|f_2\rangle$ and $|f_4\rangle$ are orthogonal to the texture-free state, and a substantial portion of the QST is stored in mode $A$, which is unaffected by radiation.
      
      When only mode $B$ undergoes Hawking radiation, mode $A$ remains unchanged, so the output state is still a product state of $A$ and the $B$ output state. For $|f_2\rangle$ and $|f_4\rangle$, the $A$ part is $|-\rangle$, which is orthogonal to the single-mode texture-free state. Hence the two-mode fidelity can be decomposed as:
      \begin{equation*}
      \begin{split}
      F(\rho_\text{out}^{(i)},f_1)&=\langle f_1|\rho_\text{out}^{(i)}|f_1\rangle\\
      &=|\langle+|-\rangle|^2\langle+|\text{Tr}_A(\rho_\text{out}^{(i)})|+\rangle\\
      &=0.
      \end{split}
      \end{equation*}
      Note that the key here is not the QST of part $B$. In fact, $|+i\rangle$ and $|-i\rangle$ themselves carry single-mode QST, and their density matrices do change under Hawking radiation. However, the orthogonality of the $A$ part to the single-mode textureless state makes the overlap of the overall state with the two-mode textureless state $f_1$ always zero, forming an "orthogonality protection barrier." Therefore, the two-mode texture measures of $|f_2\rangle$ and $|f_4\rangle$ do not change with the Hawking temperature. In contrast, the $A$ part of $|f_3\rangle$ is exactly parallel to the single-mode textureless state, with no orthogonality protection. Its two-mode QST is entirely determined by the radiated mode $B$ and therefore degrades as the Hawking temperature increases.
      
      \subsection{Only Mode $A$ Undergoes Hawking Radiation}
      In the previous subsection, we analyzed the case in which only mode $B$ undergoes Hawking radiation and found that whether the QST degrades depends on whether the mode unaffected by radiation is orthogonal to the single-mode texture-free state $|+\rangle$. A natural question is how the conclusions would change if the radiation occurs in mode $A$.
      
      Assume that only mode $A$ undergoes Hawking radiation from a Schwarzschild black hole, while mode $B$ is unaffected. The Kruskal state of mode $A$ is expanded as
      \begin{equation*}
      	\begin{split}
      		|0\rangle_K^A&=\mu|0\rangle_\text{out}^A|0\rangle_\text{in}^A+\nu|1\rangle_\text{out}^A|1\rangle_\text{in}^A,\\
      		|1\rangle_K^A&=|1\rangle_\text{out}^A|0\rangle_\text{in}^A.
      	\end{split}
      \end{equation*}
      In this setting, the computational basis is expanded in terms of the out-mode and in-mode bases as:
      \begin{equation*}
      	\begin{split}
      		&|0\rangle^A|0\rangle^B\to\mu|0\rangle_\text{out}^A|0\rangle_\text{in}^A|0\rangle^B+\nu|1\rangle_\text{out}^A|1\rangle_\text{in}^A|0\rangle^B,\\
      		&|0\rangle^A|1\rangle^B\to\mu|0\rangle_\text{out}^A|0\rangle_\text{in}^A|1\rangle^B+\nu|1\rangle_\text{out}^A|1\rangle_\text{in}^A|1\rangle^B,\\
      		&|1\rangle^A|0\rangle^B\to\mu|1\rangle_\text{out}^A|0\rangle_\text{in}^A|0\rangle^B,\\
      		&|1\rangle^A|1\rangle^B\to\mu|1\rangle_\text{out}^A|0\rangle_\text{in}^A|1\rangle^B.
      	\end{split}
      \end{equation*} 
      
      We denote by $|\Psi_k\rangle$ the joint state of the exterior and interior modes obtained after applying the Bogoliubov transformation to $|f_k\rangle$ in mode $A$. Taking the partial trace over the interior mode yields the externally accessible output state:
      \begin{equation*}
      	\rho_\text{out}^{(k)}=\text{Tr}_\text{in}(|\Psi_k\rangle\langle\Psi_k|).
      \end{equation*}
      Since $f_1$ is a pure state, the fidelity between $\rho_\text{out}^{(k)}$ and $f_1$ can be written as
      \begin{equation*}
      F(\rho_\text{out}^{(k)},f_1)=\langle f_1|\rho_\text{out}^{(k)}|f_1\rangle.
      \end{equation*}
      Using the definition of the partial trace, this can be written in a form more convenient for calculation:
      \begin{equation}
      \langle f_1|\rho_\text{out}^{(k)}|f_1\rangle=\sum_{j=0,1}|\langle f_1|\otimes\langle j|_\text{in}|\Psi_k\rangle|^2,
      \end{equation}
      where $\langle f_1|$ acts on the space $A_\text{out}\otimes B$. The interior-mode index \(j\) is eliminated in the summation. This formula combines the partial-trace and fidelity steps. In the following calculations, for convenience, we abbreviate $|i\rangle_\text{out}^A|j\rangle_\text{in}^A|k\rangle^B$ as $|ij_\text{in}k\rangle$.
      
      {\boldmath $|f_2\rangle$:} Replacing each computational basis state in $|f_2\rangle$ with its Bogoliubov expansion and rearranging, we obtain:
      \begin{equation*}
      \begin{split}
      |\Psi_2\rangle&=\frac{1}{2}[\mu|00_\text{in}0\rangle+i\mu|00_\text{in}1\rangle-|10_\text{in}0\rangle-i|10_\text{in}1\rangle\\
      &+\nu|11_\text{in}0\rangle+i\nu|11_\text{in}1\rangle].
      \end{split}
      \end{equation*}
      Compute the $j=0$ term:
      \begin{small}
      \begin{equation*}
      	\begin{split}
      		(\langle f_1|\otimes\langle0|_\text{in})|\Psi_2\rangle&=\frac{1}{2}(\mu\langle f_1|00\rangle+i\mu\langle f_1|01\rangle-\langle f_1|10\rangle-i\langle f_1|11\rangle)\\
      		&=\frac{1}{4}(\mu+i\mu-1-i)\\
      		&=\frac{1}{4}(\mu-1)(1+i).
      	\end{split}
      \end{equation*}
      \end{small}
      Compute the $j=1$ term:
      \begin{small}
      	\begin{equation*}
      		\begin{split}
      			(\langle f_1|\otimes\langle1|_\text{in})|\Psi_2\rangle&=\frac{1}{2}(\nu\langle f_1|10\rangle+i\nu\langle f_1|11\rangle)\\
      			&=\frac{\nu}{4}(1+i).
      		\end{split}
      	\end{equation*}
      \end{small}
      Therefore, we have
      \begin{equation*}
      \begin{split}
      \langle f_1|\rho_\text{out}^{(2)}|f_1\rangle&=|\frac{1}{4}(\mu-1)(1+i)|^2+|\frac{\nu}{4}(1+i)|^2\\
      &=\frac{1}{8}(\mu^2-2\mu+1+\nu^2)\\
      &=\frac{1-\mu}{4}.
      \end{split}
      \end{equation*}
      Then
      \begin{equation*}
      \mathcal{T}_F(\rho_\text{out}^{(2)})=1-\langle f_1|\rho_\text{out}^{(2)}|f_1\rangle=\frac{3+\mu}{4}.
      \end{equation*}
      
      {\boldmath $|f_3\rangle$:} Substituting the Bogoliubov transformation into the expression for $|f_3\rangle$, $|\Psi\rangle_3$ can be written as:
      \begin{equation*}
      	\begin{split}
      		|\Psi_3\rangle&=\frac{1}{2}[\mu|00_\text{in}0\rangle-\mu|00_\text{in}1\rangle+|10_\text{in}0\rangle-|10_\text{in}1\rangle\\
      		&+\nu|11_\text{in}0\rangle-\nu|11_\text{in}1\rangle].
      	\end{split}
      \end{equation*}
      Compute the $j=0$ term:
      \begin{small}
      	\begin{equation*}
      		\begin{split}
      			(\langle f_1|\otimes\langle0|_\text{in})|\Psi_3\rangle&=\frac{1}{2}(\mu\langle f_1|00\rangle-\mu\langle f_1|01\rangle+\langle f_1|10\rangle-\langle f_1|11\rangle)\\
      			&=\frac{1}{4}(\mu-\mu+1-1)\\
      			&=0.
      		\end{split}
      	\end{equation*}
      \end{small}
      Compute the $j=1$ term:
      \begin{small}
      	\begin{equation*}
      		\begin{split}
      			(\langle f_1|\otimes\langle1|_\text{in})|\Psi_3\rangle&=\frac{1}{2}(\nu\langle f_1|10\rangle-\nu\langle f_1|11\rangle)\\
      			&=0.
      		\end{split}
      	\end{equation*}
      \end{small}
      These results imply that
      \begin{equation*}
      \mathcal{T}_F(\rho_\text{out}^{(3)})=1-\langle f_1|\rho_\text{out}^{(3)}|f_1\rangle=1-0=1.
      \end{equation*}
      
      {\boldmath $|f_4\rangle$:} After calculation and simplification, for $|f_4\rangle$ we have
      \begin{equation*}
      	\mathcal{T}_F(\rho_\text{out}^{(4)})=1-\langle f_1|\rho_\text{out}^{(4)}|f_1\rangle=\frac{3+\mu}{4}.
      \end{equation*}
      
      The results for the fidelity measures of the three maximally textured resource states are summarized as follows:
      \begin{equation*}
      	\begin{split}
      		&\mathcal{T}_F(\rho_\text{out}^{(2)})=\mathcal{T}_F(\rho_\text{out}^{(4)})=\frac{3+\mu}{4},\\
      		&\mathcal{T}_F(\rho_\text{out}^{(3)})=1.
      	\end{split}
      \end{equation*}
      The corresponding texture distinguishability are:
      \begin{equation*}
      	\mathcal{D}_{24}=0,\,\mathcal{D}_{23}=\mathcal{D}_{34}=\frac{1-\mu}{4}.
      \end{equation*}
      As $T\to 0$, $\mu\to 1$, then we have
      \begin{equation*}
      	\mathcal{T}_F(\rho_\text{out}^{(2)})=\mathcal{T}_F(\rho_\text{out}^{(4)})\to1,\,\mathcal{D}_{23}=\mathcal{D}_{34}\to0.
      \end{equation*}
      This means that all maximally resourceful states retain maximal texture and are indistinguishable under the QST measure. As $T\to\infty$, $\mu\to\frac{1}{\sqrt{2}}$, then
      \begin{equation*}
      	\mathcal{T}_F(\rho_\text{out}^{(2)})=\mathcal{T}_F(\rho_\text{out}^{(4)})\to\frac{3+\frac{1}{\sqrt{2}}}{4},\,\mathcal{D}_{23}=\mathcal{D}_{34}\to\frac{1-\frac{1}{\sqrt{2}}}{4}.
      \end{equation*}
      As shown in Figure 2, when only mode $A$ undergoes Hawking radiation, the QST of $|f_2\rangle$ and $|f_4\rangle$ decreases monotonically with the Hawking temperature $T$, and their texture distinguishability from $|f_3\rangle$ increases monotonically.
      \begin{figure}[htbp]
      	\centering
      	\includegraphics[width=0.5\textwidth]{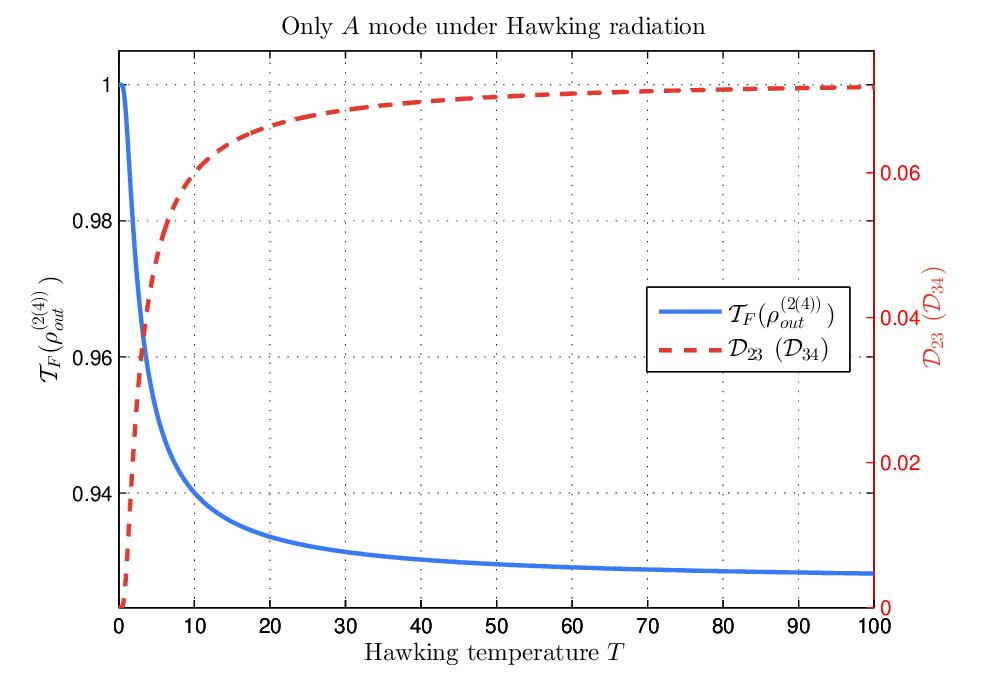}
      	\vspace{-1em} \caption{When only mode A undergoes Hawking radiation, variation of the fidelity measure $\mathcal{T}_F(\rho_\text{out}^{(2)})$($\mathcal{T}_F(\rho_\text{out}^{(4)})$) and the texture distinguishability $\mathcal{D}_{23}$ ($\mathcal{D}_{34}$) for $|f_2\rangle$ ($|f_4\rangle$) with Hawking temperature $T$. The blue solid line denotes $\mathcal{T}_F(\rho_\text{out}^{(2)})$($\mathcal{T}_F(\rho_\text{out}^{(4)})$), and the red dashed line denotes $\mathcal{D}_{23}$ ($\mathcal{D}_{34}$). $\mathcal{T}_F(\rho_\text{out}^{(2)})$($\mathcal{T}_F(\rho_\text{out}^{(4)})$) decreases monotonically from 1 to about 0.927, while $\mathcal{D}_{23}$ ($\mathcal{D}_{34}$) increases monotonically from 0 to about 0.0732.} \label{Fig.2}
      \end{figure}
      
      The results for only mode $A$ undergoing Hawking radiation are exactly opposite to those for only mode $B$. This contrast arises from the product structure of the three maximally textured resource states. When one mode undergoes radiation, if the other mode, which is unaffected by radiation, is orthogonal to the single-mode texture-free state $|+\rangle$, it provides orthogonality protection, making the overlap of the overall state with $f_1$ zero and keeping the QST measure maximal.
      
      Specifically, when mode $B$ undergoes radiation, the $A$ parts of $|f_2\rangle$ and $|f_4\rangle$ are $|-\rangle$, orthogonal to the single-mode textureless state $|+\rangle$, so the overall QST is protected. In contrast, the QST of $|f_3\rangle$ is entirely stored in mode $B$ and thus degrades as the Hawking temperature $T$ increases. Conversely, when mode $A$ undergoes radiation, the $B$ part of $|f_3\rangle$ is $|-\rangle$, orthogonal to $|+\rangle$, so the overall QST is protected; the $B$ parts of $|f_2\rangle$ and $|f_4\rangle$ have overlap $\frac{1}{2}$ with $|+\rangle$, receiving no orthogonality protection and thus degrading. In general, whether the QST degrades under the Hawking effect is determined by where the QST is stored and whether the mode unaffected by radiation provides orthogonality protection.
      
      \subsection{Both Modes $A$ and $B$ Undergo Hawking Radiation}
      In the previous two sections, we examined the cases in which only mode $B$ and only mode $A$ undergo Hawking radiation, respectively, and found that the orthogonality protection mechanism keeps the texture measure of certain maximally textured resource states unchanged. A logical and interesting question is whether orthogonality protection still holds when both modes undergo Hawking radiation simultaneously.
      
      Since modes $A$ and $B$ undergo Hawking radiation simultaneously, substituting the Bogoliubov transformation into the expansion involves four modes: $A_\text{out}$, $A_\text{in}$, $B_\text{out}$, and $B_\text{in}$. For convenience, we simplify the notation as follows:
      \begin{equation*}
      |ij_\text{in}^Akl_\text{in}^B\rangle=|i\rangle_\text{out}^A|j\rangle_\text{in}^A|k\rangle_\text{out}^B|l\rangle_\text{in}^B.
      \end{equation*}
      Replacing each of the four computational basis states with its Bogoliubov expansion:
      \begin{equation*}
      	\begin{split}
      		&|0\rangle^A|0\rangle^B\to\mu^2|00_\text{in}^A00_\text{in}^B\rangle+\mu\nu|00_\text{in}^A11_\text{in}^B\rangle+\mu\nu|11_\text{in}^A00_\text{in}^B\rangle\\
      		&~~~~~~~~~~~~~~~+\nu^2|11_\text{in}^A11_\text{in}^B\rangle,\\
      		&|0\rangle^A|1\rangle^B\to\mu|00_\text{in}^A10_\text{in}^B\rangle+\nu|11_\text{in}^A10_\text{in}^B\rangle,\\
      		&|1\rangle^A|0\rangle^B\to\mu|10_\text{in}^A00_\text{in}^B\rangle+\nu|10_\text{in}^A11_\text{in}^B\rangle,\\
      		&|1\rangle^A|1\rangle^B\to|10_\text{in}^A10_\text{in}^B\rangle.
      	\end{split}
      \end{equation*} 
      
      The joint state of the exterior and interior modes obtained after $|f_k\rangle$ undergoes Bogoliubov transformations in both modes $A$ and $B$ is denoted by $|\Psi_k\rangle$. Taking the partial trace of $|\Psi_k\rangle$ over the interior modes $A_\text{in}$ and $B_\text{in}$, we obtain the externally accessible output state:
      \begin{equation}
      \rho_\text{out}^{(k)}=\text{Tr}_{A_\text{in}B_\text{in}}(|\Psi_k\rangle\langle\Psi_k|).
      \end{equation}
      In this case, a more convenient way to compute the fidelity is:
      \begin{equation}
      \langle f_1|\rho_\text{out}^{(k)}|f_1\rangle=\sum_{j,l=0,1}|(\langle f_1|\otimes\langle j|_\text{in}^A\otimes\langle l|_\text{in}^B)|\Psi_k\rangle|^2,
      \end{equation}
      where $\langle f_1|$ acts on the subspace $A_\text{out}\otimes B_\text{out}$.
      
      {\boldmath $|f_2\rangle$:} Expanding each computational basis state in $|f_2\rangle$ via the Bogoliubov transformation and simplifying, we obtain:
      \begin{equation*}
      	\begin{split}
      		|\Psi_2\rangle&=\frac{1}{2}[\mu^2|00_\text{in}^A00_\text{in}^B\rangle+\mu\nu|00_\text{in}^A11_\text{in}^B\rangle+\mu\nu|11_\text{in}^A00_\text{in}^B\rangle\\
      		&+\nu^2|11_\text{in}^A11_\text{in}^B\rangle+i\mu|00_\text{in}^A10_\text{in}^B\rangle+i\nu|11_\text{in}^A10_\text{in}^B\rangle\\
      		&-\mu|10_\text{in}^A00_\text{in}^B\rangle-\nu|10_\text{in}^A11_\text{in}^B\rangle-i|10_\text{in}^A10_\text{in}^B\rangle].
      	\end{split}
      \end{equation*}
      Compute the $j=0, l=0$ term:
      \begin{small}
      	\begin{equation*}
      		\begin{split}
      			(\langle f_1|\otimes\langle0|_\text{in}^A\otimes\langle0|_\text{in}^B)|\Psi_2\rangle&=\frac{1}{2}(\mu^2\langle f_1|00\rangle+i\mu\langle f_1|01\rangle\\
      			&-\mu\langle f_1|10\rangle-i\langle f_1|11\rangle)\\
      			&=\frac{1}{4}(\mu^2+i\mu-\mu-1)\\
      			&=\frac{1}{4}(\mu-1)(\mu+i).
      		\end{split}
      	\end{equation*}
      \end{small}
      Compute the $j=0, l=1$ term:
      \begin{small}
      	\begin{equation*}
      		\begin{split}
      			(\langle f_1|\otimes\langle0|_\text{in}^A\otimes\langle1|_\text{in}^B)|\Psi_2\rangle&=\frac{1}{2}(\mu\nu\langle f_1|01\rangle-\nu\langle f_1|11\rangle)\\
      			&=\frac{1}{4}(\mu\nu-\nu)\\
      			&=\frac{1}{4}\nu(\mu-1).
      		\end{split}
      	\end{equation*}
      \end{small}
      Compute the $j=1, l=0$ term:
      \begin{small}
      	\begin{equation*}
      		\begin{split}
      			(\langle f_1|\otimes\langle1|_\text{in}^A\otimes\langle0|_\text{in}^B)|\Psi_2\rangle&=\frac{1}{2}(\mu\nu\langle f_1|10\rangle+i\nu\langle f_1|11\rangle)\\
      			&=\frac{1}{4}(\mu\nu+i\nu)\\
      			&=\frac{1}{4}\nu(\mu+i).
      		\end{split}
      	\end{equation*}
      \end{small}
      Compute the $j=1, l=1$ term:
      \begin{small}
      	\begin{equation*}
      		(\langle f_1|\otimes\langle1|_\text{in}^A\otimes\langle1|_\text{in}^B)|\Psi_2\rangle=\frac{1}{2}\nu^2\langle f_1|11\rangle=\frac{1}{4}\nu^2.
      	\end{equation*}
      \end{small}
      Then one has
      \begin{equation*}
      	\begin{split}
      		\langle f_1|\rho_\text{out}^{(2)}|f_1\rangle&=|\frac{1}{4}(\mu-1)(\mu+i)|^2+|\frac{1}{4}\nu(\mu-1)|^2\\
      		&+|\frac{1}{4}\nu(\mu+i)|^2+|\frac{1}{4}\nu^2|^2\\
      		&=\frac{1}{8}(\mu^2-2\mu+1+\nu^2)\\
      		&=\frac{1-\mu}{4}.
      	\end{split}
      \end{equation*}
      So the fidelity-based measure can be written as
      \begin{equation*}
      	\mathcal{T}_F(\rho_\text{out}^{(2)})=1-\langle f_1|\rho_\text{out}^{(2)}|f_1\rangle=\frac{3+\mu}{4}.
      \end{equation*}
      
      {\boldmath $|f_3\rangle$:} Replacing each computational basis state in $|f_3\rangle$ with its Bogoliubov expansion and rearranging, we can obtain that:
      \begin{equation*}
      	\begin{split}
      		|\Psi_3\rangle&=\frac{1}{2}[\mu^2|00_\text{in}^A00_\text{in}^B\rangle+\mu\nu|00_\text{in}^A11_\text{in}^B\rangle+\mu\nu|11_\text{in}^A00_\text{in}^B\rangle\\
      		&+\nu^2|11_\text{in}^A11_\text{in}^B\rangle-\mu|00_\text{in}^A10_\text{in}^B\rangle-\nu|11_\text{in}^A10_\text{in}^B\rangle\\
      		&+\mu|10_\text{in}^A00_\text{in}^B\rangle+\nu|10_\text{in}^A11_\text{in}^B\rangle-|10_\text{in}^A10_\text{in}^B\rangle].
      	\end{split}
      \end{equation*}
      Compute the $j=0, l=0$ term:
      \begin{small}
      	\begin{equation*}
      		\begin{split}
      			(\langle f_1|\otimes\langle0|_\text{in}^A\otimes\langle0|_\text{in}^B)|\Psi_3\rangle&=\frac{1}{2}(\mu^2\langle f_1|00\rangle-\mu\langle f_1|01\rangle\\
      			&+\mu\langle f_1|10\rangle-\langle f_1|11\rangle)\\
      			&=\frac{1}{4}(\mu^2-\mu+\mu-1)\\
      			&=\frac{1}{4}(\mu^2-1).
      		\end{split}
      	\end{equation*}
      \end{small}
      Compute the $j=0, l=1$ term:
      \begin{small}
      	\begin{equation*}
      		\begin{split}
      			(\langle f_1|\otimes\langle0|_\text{in}^A\otimes\langle1|_\text{in}^B)|\Psi_3\rangle&=\frac{1}{2}(\mu\nu\langle f_1|01\rangle+\nu\langle f_1|11\rangle)\\
      			&=\frac{1}{4}(\mu\nu+\nu)\\
      			&=\frac{1}{4}\nu(\mu+1).
      		\end{split}
      	\end{equation*}
      \end{small}
      Compute the $j=1, l=0$ term:
      \begin{small}
      	\begin{equation*}
      		\begin{split}
      			(\langle f_1|\otimes\langle1|_\text{in}^A\otimes\langle0|_\text{in}^B)|\Psi_3\rangle&=\frac{1}{2}(\mu\nu\langle f_1|10\rangle-\nu\langle f_1|11\rangle)\\
      			&=\frac{1}{4}(\mu\nu-\nu)\\
      			&=\frac{1}{4}\nu(\mu-1).
      		\end{split}
      	\end{equation*}
      \end{small}
      Compute the $j=1, l=1$ term:
      \begin{small}
      	\begin{equation*}
      			(\langle f_1|\otimes\langle1|_\text{in}^A\otimes\langle1|_\text{in}^B)|\Psi_3\rangle=\frac{1}{2}\nu^2\langle f_1|11\rangle=\frac{1}{4}\nu^2.
      	\end{equation*}
      \end{small}
      Therefore, we have
      \begin{equation*}
      \begin{split}
      \langle f_1|\rho_\text{out}^{(3)}|f_1\rangle&=|\frac{1}{4}(\mu^2-1)|^2+|\frac{1}{4}\nu(\mu+1)|^2\\
      &+|\frac{1}{4}\nu(\mu-1)|^2+|\frac{1}{4}\nu^2|^2\\
      &=\frac{1}{16}[(\mu^2-1)^2+\nu^2(\mu+1)^2+\nu^2(\mu-1)^2+\nu^4]\\
      &=\frac{1}{16}[\nu^4+2\nu^2(\mu^2+1)+\nu^4]\\
      &=\frac{\nu^2}{4}.
      \end{split}
      \end{equation*}
      This means that
      \begin{equation*}
      \mathcal{T}_F(\rho_\text{out}^{(3)})=1-\langle f_1|\rho_\text{out}^{(3)}|f_1\rangle=\frac{3+\mu^2}{4}.
      \end{equation*}
      
      {\boldmath $|f_4\rangle$:} Applying the Bogoliubov transformation to each computational basis state in $|f_4\rangle$ and calculating, we obtain:
      \begin{equation*}
      	\mathcal{T}_F(\rho_\text{out}^{(4)})=1-\langle f_1|\rho_\text{out}^{(4)}|f_1\rangle=\frac{3+\mu}{4}.
      \end{equation*}
      
      When both modes undergo Hawking radiation simultaneously, the fidelity measures of the three maximally textured resource states are:
      \begin{equation*}
      	\begin{split}
      		&\mathcal{T}_F(\rho_\text{out}^{(2)})=\mathcal{T}_F(\rho_\text{out}^{(4)})=\frac{3+\mu}{4},\\
      		&\mathcal{T}_F(\rho_\text{out}^{(3)})=\frac{3+\mu^2}{4}.
      	\end{split}
      \end{equation*}
      The corresponding texture distinguishability are:
      \begin{equation*}
      	\mathcal{D}_{24}=0,\,\mathcal{D}_{23}=\mathcal{D}_{34}=\frac{\mu(1-\mu)}{4}.
      \end{equation*}
      We now discuss the limiting behavior. As $T\to 0$, $\mu\to 1$, then 
      \begin{equation*}
      	\mathcal{T}_F(\rho_\text{out}^{(2)})=\mathcal{T}_F(\rho_\text{out}^{(3)})=\mathcal{T}_F(\rho_\text{out}^{(4)})\to1,\,\mathcal{D}_{23}=\mathcal{D}_{34}\to0.
      \end{equation*}
      This means that all maximally resourceful states retain maximal texture and are indistinguishable under the QST measure. As $T\to\infty$, $\mu\to\frac{1}{\sqrt{2}}$, then
      \begin{equation*}
      	\begin{split}
      	&\mathcal{T}_F(\rho_\text{out}^{(2)})=\mathcal{T}_F(\rho_\text{out}^{(4)})\to\frac{3+\frac{1}{\sqrt{2}}}{4}\approx0.927,\\
      	&\mathcal{T}_F(\rho_\text{out}^{(3)})\to\frac{3+\frac{1}{2}}{4}=0.875.\\
      	&\mathcal{D}_{23}=\mathcal{D}_{34}\to\frac{\sqrt{2}-1}{8}\approx0.0518.
      	\end{split}
      \end{equation*}
      
      As shown in Figure 3, when both modes undergo Hawking radiation simultaneously, both $\mathcal{T}_F(\rho_\text{out}^{(2(4))})$ and $\mathcal{T}_F(\rho_\text{out}^{(3)})$ decrease monotonically with increasing Hawking temperature.
      \begin{figure}[htbp]
      	\centering
      	\includegraphics[width=0.5\textwidth]{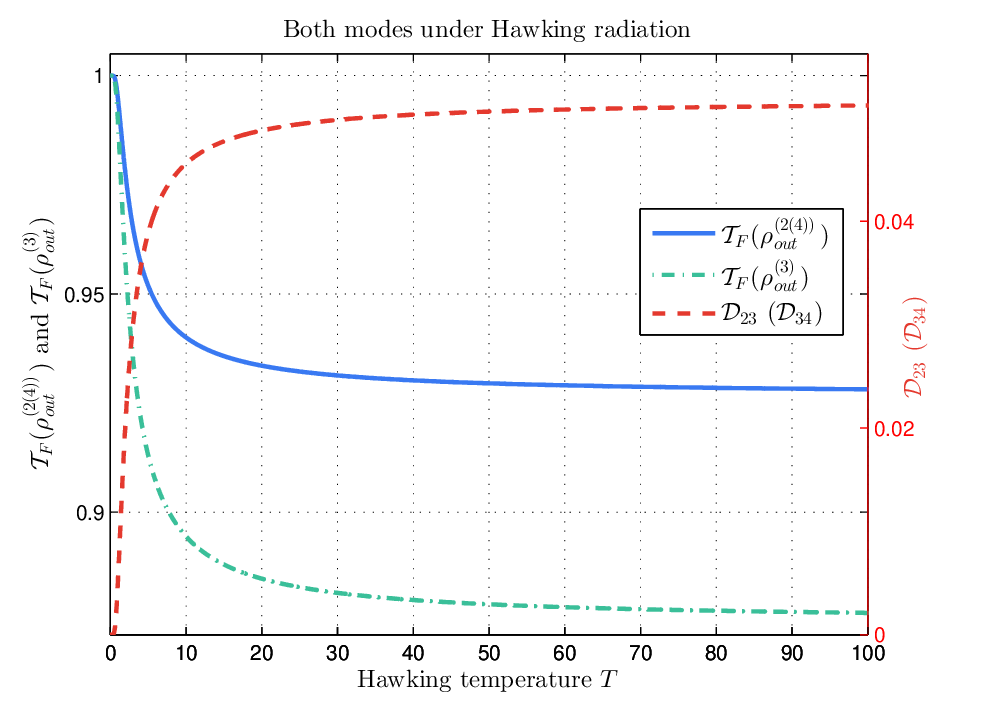}
      	\vspace{-1em} \caption{When both modes undergo Hawking radiation simultaneously, $\mathcal{T}_F(\rho_\text{out}^{(2(4))})$ (blue solid line), $\mathcal{T}_F(\rho_\text{out}^{(3)})$ (green dash-dotted line), and $\mathcal{D}_{23}$ ($\mathcal{D}_{34}$) (red dashed line) vary with the Hawking temperature $T$. $\mathcal{T}_F(\rho_\text{out}^{(2(4))})$ and $\mathcal{T}_F(\rho_\text{out}^{(3)})$ decrease monotonically from 1 to about 0.927 and 0.875, respectively.} \label{Fig.3}
      \end{figure}
      The QST of all three states degrades as the Hawking temperature increases, and none retains maximal texture. This contrasts with the single-mode radiation case, where one state is always orthogonality-protected: when both modes undergo radiation simultaneously, the orthogonality protection vanishes completely. The degradation of $|f_2\rangle$ and $|f_4\rangle$ is exactly the same as when only $A$ undergoes radiation. The reason is that their $B$ parts are $|+i\rangle$ and $|-i\rangle$, respectively, whose overlap with the single-mode textureless state under Hawking radiation remains $\frac{1}{2}$, exactly the same as before radiation. Therefore, the degree of QST degradation of these two states is entirely determined by mode $A$, and whether mode $B$ undergoes radiation does not change the final numerical results.
      
      {\bf Significantly, the case of $|f_3\rangle$ merits separate discussion.} Its QST is initially stored entirely in mode $B$, while its $A$ part is the single-mode texture-free state in the Kruskal basis. Hawking radiation, however, changes the basis: after the Bogoliubov transformation, $|+\rangle$ is no longer texture-free in the out basis, and its overlap with the single-mode texture-free state becomes $\frac{1+\mu}{2}$. When both modes undergo radiation simultaneously, this newly generated QST in mode $A$ partially compensates for the degradation of the QST in mode $B$. Therefore:
      \begin{equation}
      \mathcal{T}_F(\rho_\text{out}^{(3),AB})=\frac{3+\mu^2}{4}\geq\frac{1+\mu}{2}=\mathcal{T}_F(\rho_\text{out}^{(3),B}).
      \end{equation}
      In other words, under two-mode radiation, the QST measure of $|f_3\rangle$ is actually higher than in the case where only mode $B$ undergoes radiation. This appears to contradict the intuition that "more radiation leads to more degradation." The underlying reason is that under Hawking radiation, QST can be redistributed among modes, so QST gained by one mode under radiation can partially compensate for QST lost by another.
      
      Combining the three cases yields the following conclusions: The QST degradation of maximally textured resource states under Hawking radiation depends not only on the number of radiated modes but also on the initial distribution of QST among the modes and on how each mode's overlap with the single-mode textureless state changes under the Bogoliubov transformation. Orthogonality protection can completely prevent degradation, but once both modes undergo radiation, the protection vanishes and all states suffer QST degradation. In the case of $|f_3\rangle$, the originally textureless mode $A$ acquires QST in the out basis, which instead mitigates the overall degradation.
      
      \subsection{Resource-Theoretic Interpretation of Orthogonality Protection}
      In the preceding sections, we explained orthogonality protection using product structure and orthogonality. However, this explanation depends on the specific form of the fidelity-based measure. A natural question is whether orthogonality protection is measure-dependent. That is, whether the conclusions still hold for other QST measures. In this subsection, starting from the general framework of QST resource theory, we show that orthogonality protection is not an accidental consequence of the fidelity-based measure, but a general property determined by the structure of the maximally resourceful equivalence class.
      
      In QST resource theory, free operations are CPTP maps that leave the texture-free state $f_1$ invariant. Denote the orthogonal complement of $f_1$ by
      \begin{equation}
      \mathcal{H}^{'}=\{X|\,\,\text{Tr}(f_1X)=0\}.
      \end{equation}
      It can be shown that all states in $\mathcal{H}^{'}$ belong to the same equivalence class under free operations. The construction is as follows: for any $\tau\in \mathcal{H}^{'}$, define a map \cite{jadl}
      \begin{equation}
      \Gamma_\tau(X)=\text{Tr}(f_1X)f_1+\text{Tr}[(I-f_1)X]\tau.
      \end{equation}
      This map is CPTP and satisfies $\Gamma_\tau(f_1)=f_1$, so it is a free operation. For any $\rho\in \mathcal{H}^{'}$, $\Gamma_\tau(\rho)=\tau$. Therefore, any two states in $\rho\in \mathcal{H}^{'}$ can be interconverted by free operations and belong to the same resource equivalence class.
      
      A direct consequence of the above equivalence class is that any QST measure $\mathcal{T}$ that is monotone under free operations must be constant on $\mathcal{H}^{'}$. The proof is simple: if $\rho$, $\sigma\in \mathcal{H}^{'}$ can be interconverted by free operations, monotonicity gives $\mathcal{T}(\rho)\leq \mathcal{T}(\sigma)$ and $\mathcal{T}(\sigma)\leq \mathcal{T}(\rho)$, so $\mathcal{T}(\rho)= \mathcal{T}(\sigma)$. Since the Fourier states $|f_k\rangle$ ($k\geq 2$) lie in $\mathcal{H}^{'}$, and any valid QST measure attains its maximum on the Fourier states, this constant value of $\mathcal{T}$ on $\mathcal{H}^{'}$ is its maximum. This conclusion does not depend on the specific form of the measure; it requires only monotonicity, a basic condition for a QST measure.
      
      Now we apply the above conclusions to the physical problem considered in this work. When only mode $A$ or only mode $B$ undergoes Hawking radiation, the other mode remains unchanged, and the output state is still a product state, whose overlap with $f_1$ can be factorized as:
      \begin{equation}
      \langle f_1|\rho_\text{out}|f_1\rangle=|\langle+|\phi_\text{free}\rangle|^2.\langle+|\rho_\text{out}^\text{rad}|+\rangle,
      \end{equation}
      where $|\phi_\text{free}\rangle$ denotes the state of the unaffected mode and $\rho_\text{out}^\text{rad}$ the reduced density matrix of the radiated mode.
      
      $\bullet$ When only mode $B$ undergoes radiation, the $A$ parts of $|f_2\rangle$ and $|f_4\rangle$ are $|-\rangle$, which is orthogonal to the single-mode texture-free state $|+\rangle$; the first factor above is zero, so the output state lies in $\mathcal{H}^{'}$. By the general conclusion from resource theory, their QST measures must attain the maximum, independent of the choice of measure and the Hawking temperature. The $A$ part of $|f_3\rangle$ is $|+\rangle$, so the factor is 1; the overall output state leaves $\mathcal{H}^{'}$, and the QST degrades.
      
      $\bullet$ When only mode $A$ undergoes radiation: the $B$ part of $|f_3\rangle$ is $|-\rangle$, orthogonal to $|+\rangle$, so the overall output state lies in $\mathcal{H}^{'}$ and the QST measure retains its maximum. The $B$ parts of $|f_2\rangle$ and $|f_4\rangle$ are $|+i\rangle$ and $|-i\rangle$, respectively; the factor is nonzero, so the overall output state leaves $\mathcal{H}^{'}$ and the QST degrades.
      
      Therefore, orthogonality protection is not a separate physical mechanism; rather, its essence is the preservation of the $\mathcal{H}^{'}$ equivalence class under specific radiation conditions.
      
	\section{IV. Conclusions and discussions}
    We study the QST behavior of three maximally resourceful Fourier states $|f_2\rangle$, $|f_3\rangle$, and $|f_4\rangle$ in a two-mode Dirac field under Hawking radiation from a Schwarzschild black hole. Using the fidelity-based measure $\mathcal{T}_F$, we consider three cases: only mode $B$ undergoes radiation, only mode $A$ undergoes radiation, and both modes undergo radiation simultaneously. Our results show that when only one mode undergoes radiation, if the unaffected mode is orthogonal to the single-mode texture-free state $|+\rangle$, the overlap between the overall output state and the two-mode texture-free state $f_1$ is always zero, and the QST measure remains maximal, independent of the Hawking temperature. We call this phenomenon orthogonality protection. Specifically, when only mode $B$ undergoes radiation, the $A$ parts of $|f_2\rangle$ and $|f_4\rangle$ are $|-\rangle$, orthogonal to $|+\rangle$, so their measures remain $1$; the $A$ part of $|f_3\rangle$ is $|+\rangle$, which provides no orthogonality protection, and its measure degrades to $\frac{1+\mu}{2}$. When only mode $A$ undergoes radiation, the conclusion is symmetric: $|f_3\rangle$ is protected, while $|f_2\rangle$ and $|f_4\rangle$ degrade to $\frac{3+\mu}{4}$. When both modes undergo radiation simultaneously, orthogonality protection disappears completely, and the texture measures of all three Fourier states decrease with increasing Hawking temperature. In particular, the QST measure of $|f_3\rangle$ under two-mode radiation, $\frac{3+\mu^2}{4}$, is higher than $\frac{1+\mu}{2}$ under only mode $B$ radiation. This result seems to contradict the intuition that more radiation leads to more degradation. The physical reason is that the $A$ part of $|f_3\rangle$ is textureless in the Kruskal basis, but after the Bogoliubov transformation it is no longer textureless in the out basis, and its overlap with the single-mode textureless state becomes $\frac{1+\mu}{2}$. Thus, the newly acquired QST of the originally textureless mode $A$ under radiation partially compensates for the degradation of mode $B$, indicating that QST can be redistributed among modes.
    
    From a resource-theoretic perspective, orthogonality protection admits a unified explanation. All states in the orthogonal complement $\mathcal{H}^{'}$ of the textureless state $f_1$ belong to the same equivalence class under free operations. Therefore, any QST measure that is monotonic under free operations is constant on $\mathcal{H}^{'}$, and this constant equals the maximum. When only one mode undergoes radiation and the unaffected mode is orthogonal to $|+\rangle$, the overall output state always remains in $\mathcal{H}^{'}$, so the QST measure remains maximal, independent of the choice of measure and the Hawking temperature. Once both modes undergo radiation simultaneously, the output state leaves $\mathcal{H}^{'}$, and the measures degrade for all states. This conclusion does not depend on the specific form of the measure; it requires only monotonicity.
    
    Unlike entanglement and coherence, which typically degrade monotonically with radiation, QST involves all elements of the density matrix and can exhibit nonmonotonic behavior such as orthogonality protection and intermode compensation. This work considers only the two-mode, $D=4$ case. Future work could extend this to more modes, higher-dimensional systems, and black hole backgrounds such as Reissner–Nordström and Kerr.

	\bigskip
	{\bf Acknowledgments:} ~This work is supported by the National Natural Science Foundation of China (NSFC) under Grant No. 12171044 and the specific research fund of the Innovation Platform for Academicians of Hainan Province.

	\appendix

\end{document}